\documentclass[english]{article}
\usepackage[T1]{fontenc}
\usepackage[latin9]{inputenc}
\usepackage{geometry}
\usepackage{fancyhdr}
\usepackage{array}
\usepackage{float}
\usepackage{textcomp}
\usepackage{multirow}
\usepackage{amstext}
\usepackage{amssymb}
\usepackage{graphicx}

\makeatletter

\DeclareRobustCommand{\greektext}{%
  \fontencoding{LGR}\selectfont\def\encodingdefault{LGR}}
\DeclareRobustCommand{\textgreek}[1]{\leavevmode{\greektext #1}}
\DeclareFontEncoding{LGR}{}{}
\DeclareTextSymbol{\~}{LGR}{126}
\newcommand{\lyxmathsym}[1]{\ifmmode\begingroup\def\b@ld{bold}
  \text{\ifx\math@version\b@ld\bfseries\fi#1}\endgroup\else#1\fi}

\providecommand{\tabularnewline}{\\}

\newcommand{\lyxaddress}[1]{
\par {\raggedright #1
\vspace{1.4em}
\noindent\par}
}

\date{July 18, 2014}

\makeatother

\usepackage{babel}
\begin{document}

\title{{\Huge Derivation of chemical abundances in star-forming galaxies
at intermediate redshift}}

\maketitle
\noindent \begin{center}
\vspace{0.02\paperheight}
By
\par\end{center}

\vspace{0.003\paperheight}

\noindent \begin{center}
{\Large Jose Manuel Pérez Martínez}
\par\end{center}{\Large \par}

\vspace{0.04\paperheight}

\begin{center}
UNIVERSIDAD AUTÓNOMA DE MADRID
\par\end{center}

\vspace{0.005\paperheight}

\begin{center}
Department of Theoretical Physics
\par\end{center}

\begin{center}
Master Degree in Theoretical Physics and Astrophysics
\par\end{center}

\vspace{0.02\paperheight}

\begin{center}
\includegraphics[scale=0.6]{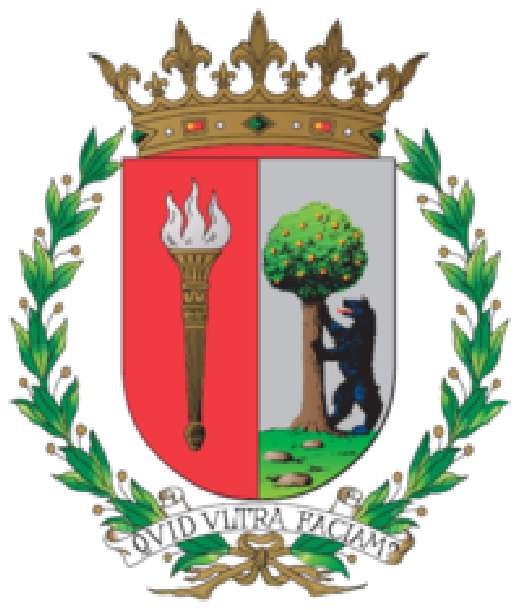}
\par\end{center}

\vspace{0.03\paperheight}

\lyxaddress{\begin{center}
{\Large Supervised by}
\par\end{center}}
\begin{quotation}
\begin{center}
{\large Ángeles Díaz Beltrán \& Carlos Hoyos Fernández de Córdova}
\par\end{center}{\large \par}
\end{quotation}
\pagenumbering{gobble}

\pagebreak{}

\pagebreak{}
\begin{abstract}
{\large We have studied a sample of 11 blue, luminous, metal-poor
galaxies at redshift $0.744<z<0.835$ from the DEEP2 redshift survey.
They were selected by the presence of the {[}OIII{]} $\lambda4363$
auroral line and the {[}OII{]} $\lambda\lambda3726,3729$ doublet
together with the strong emission nebular {[}OIII{]} lines in their
spectra from a sample of $\sim6000$ galaxies within a narrow redshift
range. All the spectra have been taken with DEIMOS, which is a multi-slit,
double-beam spectrograph which uses slitmasks to allow the spectra
from many objects (between 50 and 100 per barrel) to be imaged at
the same time. The selected objects present high luminosities ($-20.3<M_{B}<-18.5$),
remarkable blue color index, and total oxygen abundances between $7.69$
and $8.15$ which represent 1/3 to 1/10 of the solar value. The wide
spectral coverage (from $6500$ to $9100$ Å) of the DEIMOS spectrograph
and its high spectral resolution, around ($R$ around $5000$), bring
us an opportunity to study the behaviour of these star-forming galaxies
at intermediate redshift with high quality spectra. }{\large \par}

{\large We put in context our results together with others presented
in the literature up to date to try to understand the luminosity-metallicity
relation this kind of objects define. The star-forming metal-poor
galaxies would be of special relevance in showing the diversity among
galaxies of similar luminosities and could serve to understand the
processes of galaxy evolution. }{\large \par}

\vspace{0.04\paperheight}

\end{abstract}
\pagenumbering{gobble}\tableofcontents{}

\newpage{}

\pagenumbering{arabic}

\section{{\LARGE Introduction}}

{\large The chemical evolution of galaxies constitutes the cornerstone
to understand the star formation history of galaxies and it is closely
related with dynamical galaxy processes such as gas outflows from
supernovae and inflows from cosmic accretion. The mean method to measure
the chemical abundance of galaxies bases on the analysis of nebular
emission lines. Most of these emission lines can be observed in the
optical and near infra-red up to redshift of $\thicksim3$ with current
ground and space observatories, but it is not that easy to collect
the necessary data to study their properties due to instrument limitations
to cover wavelength ranges sufficiently wide.}{\large \par}

{\large Measuring the flux ratio of the {[}OIII{]}$\lambda4363$ auroral
line against the stronger nebular lines such as {[}OIII{]} $\lambda\lambda4959,5007$
ones will provides one of the most confident ways of determining metal
abundances in galaxies. However the detection of the {[}OIII{]}$\lambda4363$
line is difficult since it is very weak in the case of metal-poor
galaxies and almost indetectable in metal-rich galaxies. Consequently
the detection of {[}OIII{]} $\lambda4363$ is a strong indication
of the metal poor content of that object. One galaxy type which commonly
presents this feature is the so called blue compact galaxies (BCGs)
which were first indentified by Zwicky (1965) as faint emission-line
galaxies with UV excess. There exist a considerable interest in the
study of low-metallicity galaxies considering that they can enlighten
the not well understood first stages of galaxy formation and chemical
enrichment (Kunth et al. 2000). }{\large \par}

{\large Considerable knowledge about this kind of galaxies came thanks
to the images provided by the Hubble Space Telescope (HST) which allow
to analyze the different types of objects which form the BCGs: starburst
galaxies (Cowie et al. 1991), low luminosity dwarfs at low redshift
(Im et al. 1995), low luminosity active galactic nuclei (Tresse et
al. 1996) and compact narrow emission-line galaxies (Koo et al. 1994,
1995; Guzmán et al. 1996).}{\large \par}

{\large Within galaxies of this kind are the luminous compact blue
galaxies (LCBGs) which define those luminous ($M_{B}<-17.5$), compact
($\lyxmathsym{\textgreek{m}}_{B}\leq21.0$$mag$$arcseg\text{\textsuperscript{-}\texttwosuperior}$)
and blue ($B-V\leq0.6$) objects undergoing a major star-forming process.
In this category are included some starburst galaxies at intermediate
redshift and the most luminous BCGs (Hoyos et al. 2004). }{\large \par}

{\large The least metal-content galaxies known are I Zw 18 (Searle
\& Sargent 1972) and SBS0335-052 (Izotov et al. 1990) which have total
oxygen abundances of $12+\log(O/H)=7.14$ (Izotov et al. 2006b) and
$7.19-7.34$ respectively (Izotov et al. 1990). There have been attempts
to rise the metal-poor number of galaxies observed but up to date
there are only $~70$ known galaxies with $12+\log(O/H)\leq7.65$,
most of them in the local universe $z\lesssim1$.}{\large \par}

{\large There are not many observations of metal-poor galaxies at
intermediate redshifts (Kobulnicky et al. 2003; Kewley et al. 2004;
Hoyos et al. 2005; Kakuzu et al 2007, Ly et al 2014) which suggests
that at a given metallicity, galaxies were typically more luminous
in the past, while the high redshift samples show subsolar metallicities
with lumininosities which exceed the bright of comparable matallicity
local galaxies. Although it is expected that the luminosity-metallicity
relation evolves with the lifetime of galaxies, the way it does is
not well understood. Therefore it is needed to increase the number
of metal-poor galaxies studied in order to determine the causes of
the chemical abundances behaviour through the galaxy life-time.}{\large \par}

{\large In this work I focus in the galaxies contained in the fourth
release of DEEP2 galaxy survey with detectable {[}OIII{]} $\lambda4363$
line, with the aim of a) deriving their chemical abundances following
the so called ``direct method''; and b) studying the metallicty-luminosity
relation for galaxies at intermediate redshift ($z\sim0.7$ to $0.8$).
The studied galaxies have been selected by the presence of {[}OIII{]}
$\lambda4363$ weak emission line in their spectra together with other
oxygen lines such as {[}OIII{]} $\lambda\lambda4959,5007$ and a resolved
{[}OII{]} $\lambda\lambda3726,3729$ doublet. These conditions can
only be reached with the aid of great deep field telescopes such as
the 10m. Keck telescopes in Hawaii.}{\large \par}

{\large The outline of the work is as follows. In Section 2, we describe
the characteristics of DEIMOS spectrograph and the features and quality
of DEEP project spectra. Here is where we present the sample selection
process and conditions along with other properties collected such
as magnitudes or logarithm mass. We put in context all this information
with other galaxies at comparable redshifts of DEEP2 through color-magitude
diagrams. We then present in Section 3 our results on the measurement
of the emission lines, the derivation of the gaseous electron density
and temperature and the computation of ionic and total abundances.
In Section 4 we discuss the results obtained constructing a luminosity-metallicity
diagrams and studying the ionic oxygen abundance balance versus the
metallicity of the sample. In order to explore any possible effects
on the derived abundances of the degree of excitation of the nebulae
we also compare the data from our sample with previous works on the
same type of galaxies. Finally in Section 5 we present the main conclusions
of this study.}{\large \par}

{\large Throughout this work we assume a flat cosmology with $\lyxmathsym{\textgreek{W}}_{\lyxmathsym{\textgreek{L}}}=0.7$,
$\lyxmathsym{\textgreek{W}}_{M}=0.3$ and $H_{0}=70$ $\mbox{\ensuremath{km}}s^{-1}$
$Mpc^{-1}$ to determine distance-dependent measurements. Magnitudes
are in the Vega system.}{\large \par}

\section{{\LARGE Data Selection}}

\subsection{{\Large The DEEP project}}

{\large The Deep Extragalactic Evolutionary Probe (DEEP) is a multi-year
program which uses the twin 10-m W.M. Keck Telescopes and the Hubble
Space Telescope (HST) to conduct a large-scale survey of distant,
faint, field galaxies. The DEEP2 redshift survey is the second stage
of DEEP project which use the DEIMOS spectrograph to obtain spectra
of $\sim50.000$ faint galaxies with redshifts $z\gtrsim0.7$. DEIMOS
is a multi-slit, double-beam spectrograph which uses slitmasks to
allow the spectra from many objects (between 50 and 100 per barrel)
to be imaged onto a mosaic CCD array. Some of the most important features
are its wide spectral coverage (from 6500 to 9100 Å) and its high
spectral resolution, around $R\sim5000$.}{\large \par}

{\large The survey has been designed to have the reliability of local
redshift surveys and to be complementary to others large redshift
surveys such as the SDSS project and the 2dF survey. DEEP2 survey
observes at four separate sky fields covering approximately three
square degrees. Along this area it has detected 52989 galaxies which
form the redshift catalog; for each of them the survey provides a
one dimensional spectrum split in two images corresponding to the
red and blue beam of the instrument. The spectra are grouped according
to slitmask number and each slitmask was observed for approximately
1 hour (3 exposures of 1200 seconds each). For the data presented
in this work the 1200 lines per mm grating centered at $7800\mathring{A}$
was used, typically covering a wavelength range of $6500-9100\mathring{A}$
depending on slit placement on the mask. The dispersion obtained under
these conditions is about 3 pixels per angstrom, which allows deconvolution
of usually blended lines such as the {[}OII{]} $\lambda\lambda3726,3729$
doublet.}{\large \par}

\subsection{{\Large Sample selection}}

{\large Galaxies were selected by inspection of reduced spectra from
the DEEP2 redshift survey using several criteria. A selection process
over the entire reshift catalog was carried out in order to isolate
the fraction of spectra which contains the strong oxygen emission
lines {[}OII{]}$\lambda3727$, {[}OIII{]}$\lambda4959$, {[}OIII{]}$\lambda5007$
and the weak auroral line {[}OIII{]}$\lambda4363$. The redshift catalog
includes 52989 objects whose spectra cover in average the wavelength
range between $6500-9100\mathring{A}$. This means that only a small
portion of the survey will have the entire set of lines due to constraints
imposed by the redshift value. We can determine the correct redshift
range using the accurate redshift measurements provided by DEEP2 along
with the spectrum wavelength range mentioned before. In order to get
the constraints we need to include the two furthest lines into this
range, so that the observed {[}OII{]} $\lambda3727$ line wavelength
was longer than $6500\mathring{A}$ while the observed {[}OIII{]}
$\lambda4959$ line wavelength was shorter than $9100\mathring{A}$.
There is no need to include the {[}OIII{]}$\lambda5007$ in the range
since its flux value can easily be computed from the {[}OIII{]} $\lambda4959$
using the theoretical ratio between both lines of 2.98 (Storey and
Zeippen 2000).}{\large \par}

{\large Thus, the first data sample ready to analyze includes only
6241 galaxies within a redshift range of $0.744<z<0.835$. After this
selection process, it was necessary to carry out a visual inspection
of each spectrum setting two additional conditions: Firstly we had
to determine whether the auroral {[}OIII{]}$\lambda4363$ line is
visible in the spectrum or not. This line is crucial to understand
the properties of the galaxy such as luminosity, metal-richness and
ionization degree, moreover it could bring information about a possible
underlying population inside the galaxy (C.Hoyos and A.I. Díaz 2006).
The second step is to look for an {[}OII{]} $\lambda3727$ doublet
clearly separeted in their two energy levels $\lambda3726$ and $\lambda3729$
which will allow us to determine the electron density of the ionized
region without making any assumption in contrast with precedent works.}{\large \par}

{\large I used the IGI tool of the STSDAS package of Iraf to plot
the spectrum in three panels, each of them has at least an oxygen
emission line in the center of the chosen wavelength range. This method
allows us to perform a rapid visual inspection of each spectrum determining
the quality of its line fluxes at first sight. In the case in which
we were not capable of seeing the {[}OIII{]}$\lambda4363$ line due
to the high noise level noise we discarded the spectrum.}{\large \par}

\begin{center}

\begin{table}
\begin{centering}
\begin{tabular}{ccccccccc}
\hline 
$DEEP2$ & \multirow{2}{*}{$z$} & $R.A.$ & $Decl.$ & \multirow{2}{*}{$m_{B}$} & \multirow{2}{*}{$M_{B}$} & \multirow{2}{*}{$U-B$} & \multirow{2}{*}{$B-V$} & \multirow{2}{*}{$M_{log}^{(a)}$}\tabularnewline
$ID$ &  & $(J2000)$ & $(J2000)$ &  &  &  &  & \tabularnewline
\hline 
\hline 
13016475 & 0.74684 & 14:20:57.85 & 52:56:41.81 & 22.30 & -19.62 & 0.03 & 0.01 & 9.26\tabularnewline
22032252 & 0.74872 & 16:53:03.49 & 34:58:48.95 & 23.29 & -18.55 & 0.33 & 0.30 & 9.35\tabularnewline
31019555 & 0.75523 & 23:27:20.37 & 00:05:54.76 & 23.15 & -18.90 & -0.30 & -0.36 & 8.34\tabularnewline
12012181 & 0.77166 & 14:17:54.62 & 52:30:58.42 & 22.84 & -19.33 & 0.01 & -0.02 & 9.09\tabularnewline
14018918 & 0.77091 & 14:21:45.41 & 53:23:52.70 & 22.32 & -19.62 & 0.17 & 0.13 & 9.48\tabularnewline
41059446 & 0.77439 & 02:26:21.48 & 00:48:06.81 & 21.78 & -20.27 & 0.42 & 0.34 & 10.10\tabularnewline
41006773 & 0.78384 & 02:27:48.87 & 00:24:40.08 & 23.47 & -18.82 & 0.08 & 0.03 & 8.99\tabularnewline
22021909 & 0.79799 & 16:50:55.34 & 34:53:29.88 & 23.38 & -18.90 & 0.25 & 0.17 & 9.26\tabularnewline
22020856 & 0.79448 & 16:51:31.47 & 34:53:15.96 & 22.59 & -19.46 & 0.32 & 0.23 & 9.59\tabularnewline
22020749 & 0.79679 & 16:51:35.22 & 34:53:39.48 & 22.59 & -19.50 & 0.38 & 0.28 & 9.69\tabularnewline
31046514 & 0.78856 & 23:27:07.50 & 00:17:41.50 & 22.67 & -19.30 & 0.61 & 0.47 & 9.94\tabularnewline
\hline 
\end{tabular}
\par\end{centering}

\centering{}\caption{Right ascension, declination, redshift, DEEP2 photometric magnitudes
and logarithmic mass of the galaxy sample. All the quantities has
been obtained photometrically by blabla}
\end{table}

\end{center}

\begin{figure}[H]
\begin{center}

\includegraphics[scale=0.75]{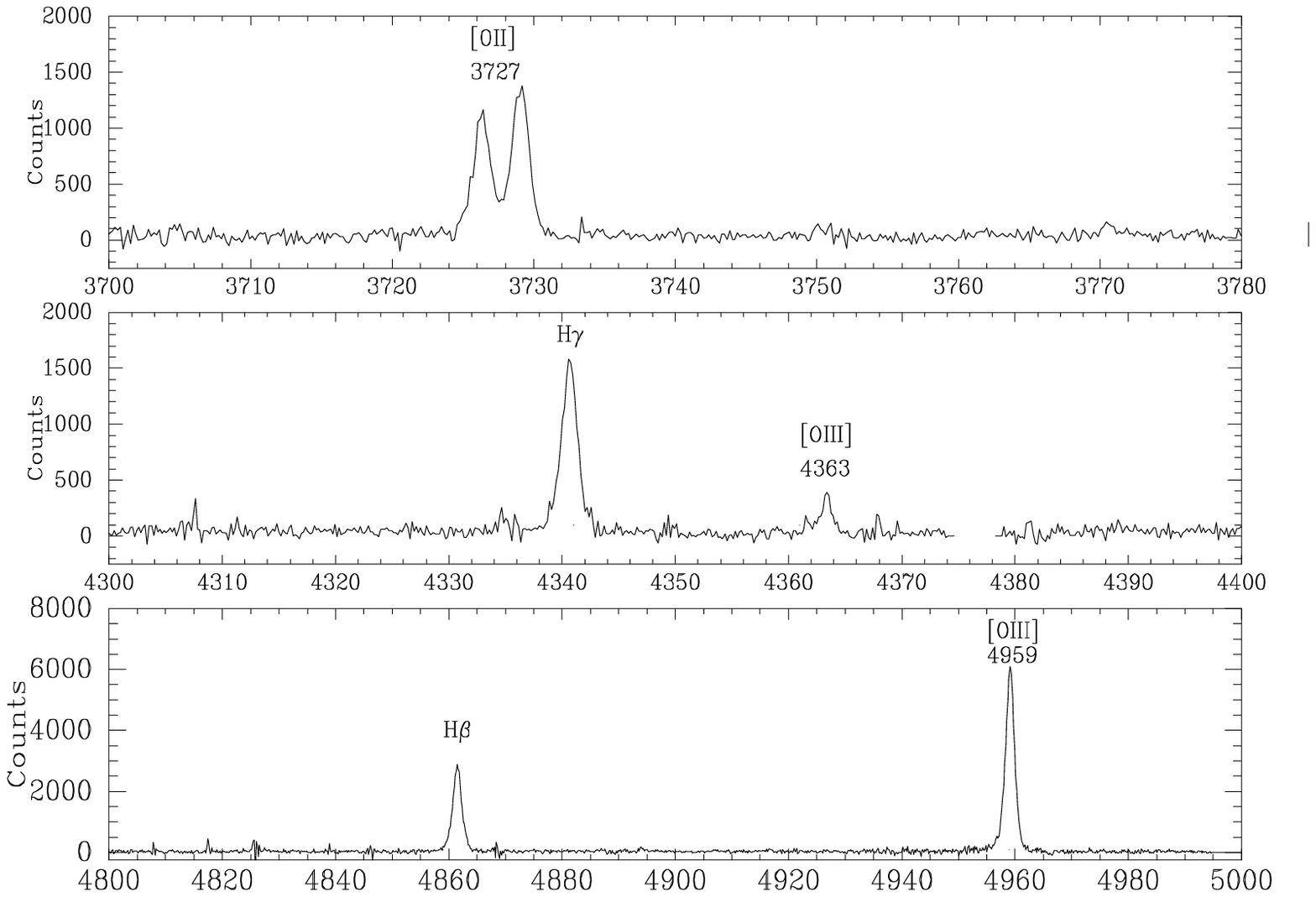}

\bigskip

\bigskip

\includegraphics[scale=0.75]{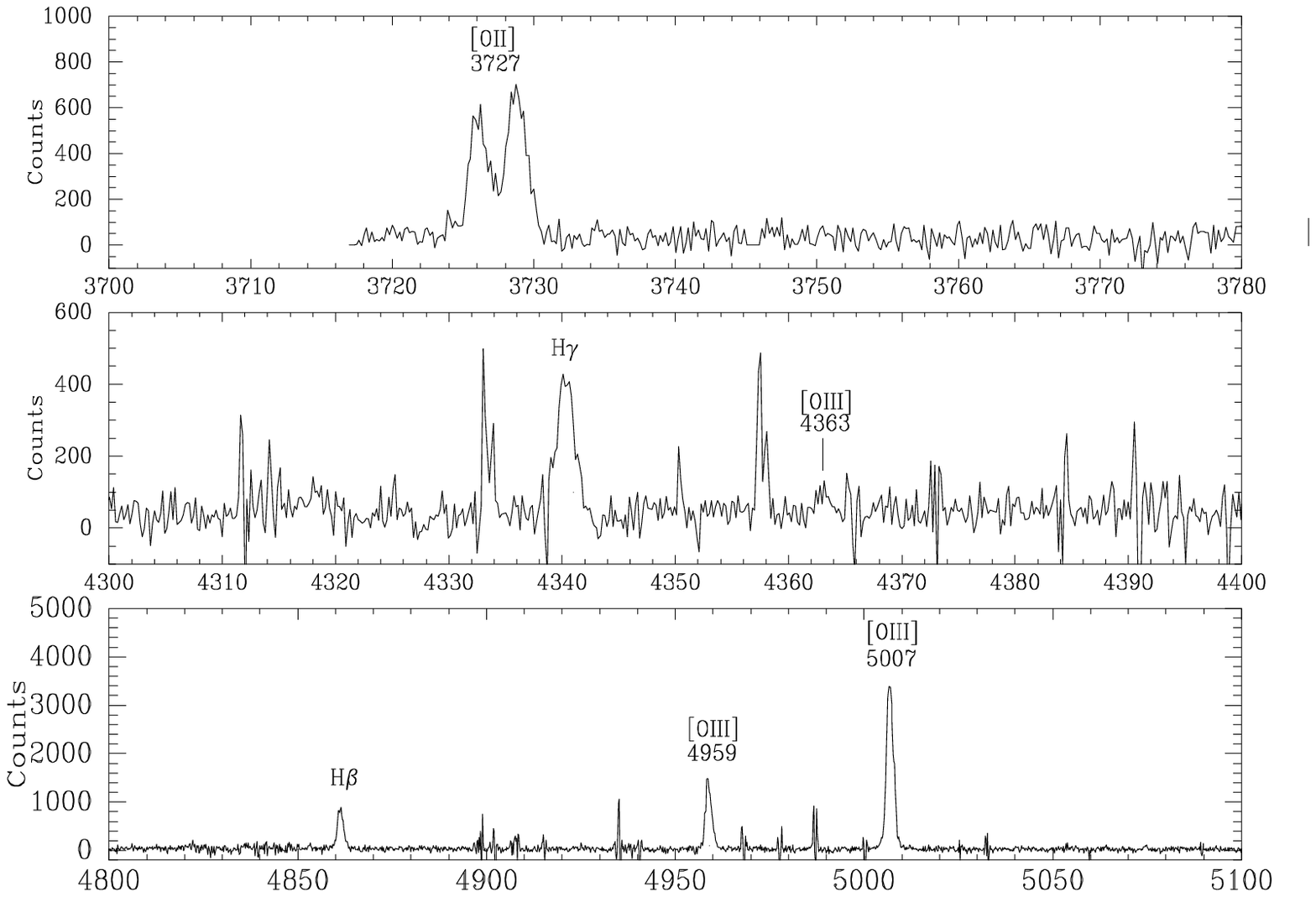}

\caption{Spectrum of two low metallicity galaxies at redshift z=0.747 and z=0.789
respectively. Both of them show the temperature sensitive {[}OIII{]}$\lambda4363$
line used to identify the sample and the other oxygen emission lines
used to measure the ionic and total abundances. Top panels do not
present the {[}OIII{]}$\lambda5007$ due to a slit displacement on
the mask which reduced the long wavelength limit in which DEIMOS can
measure.}

\end{center}
\end{figure}

{\large The best and worst quality spectra in the final sample are
represented in top and down panels of Figure 1 respectivey. As we
can see the flux intensity of the Balmer lines is somehow related
with the {[}OIII{]} flux intensity lines, so that when we find a low
signal in $H\lyxmathsym{\textgreek{g}}$ we expect to have problems
measuring the auroral {[}OIII{]}$\lambda4363$ line considering the
average noise level of spectra. This will be the greatest difficulty
in selecting the final galaxy sample and one of the major error sources
in determining the electronic temperatures and oxygen abundances.
Finally the visual inspection of the subsample yielded 11 galaxies,
wich represent the $0.2$\% of the initial sample. }{\large \par}

{\large The next step is to characterize the final sample of spectroscopically
selected galaxies. We present the photometrically observed and absolute
magnitudes for the Johnson B filter in addition to the color index
for each object included in the final sample in Table 1. Blue absolute
magnitudes ($M_{B}$) and rest frame $U-B$ and $B-V$ colors were
calculated from the BRI photometry (Coil et al 2004) with K-corrections
following those described by Willmer et al. (2005). The results show
that most of the galaxies have $M_{B}\leq18.5$ with color indexes
such as $U-B\lesssim0.6$ which means we are working with luminous
blue galaxies that are undergoing a burst of star formation (Hoyos
et al. 2004).}{\large \par}

{\large Our final galaxy sample has to be put in context within others
galaxies at similar redshift. We have chosen a set of galaxies from
the DEEP2 survey which contains 2421 objects at aproximately the same
redshift of our work. The size of this DEEP2 sample has been determined
selecting only those galaxies whose $EW_{\beta}>10\mathring{A}$.
The equivalent width has been obtained from the DEEP2 Automated Line
Fitting in 1-D Spectra (Weiner B. 2004). Linefit is a Fortran program
designed to make automatic fitting of spectral lines of 1-d spectra,
specifically DEEP2 spectra. For each spectrum, it produces an output
of line parameters including the EW of every fitted line. This procedure
is automated and it works well on emission lines. The results of the
comparison between the objects of this work and the Linefit sample
can be seen in Figure 2.}{\large \par}

\begin{figure}[H]
\begin{center}

\includegraphics[scale=0.45]{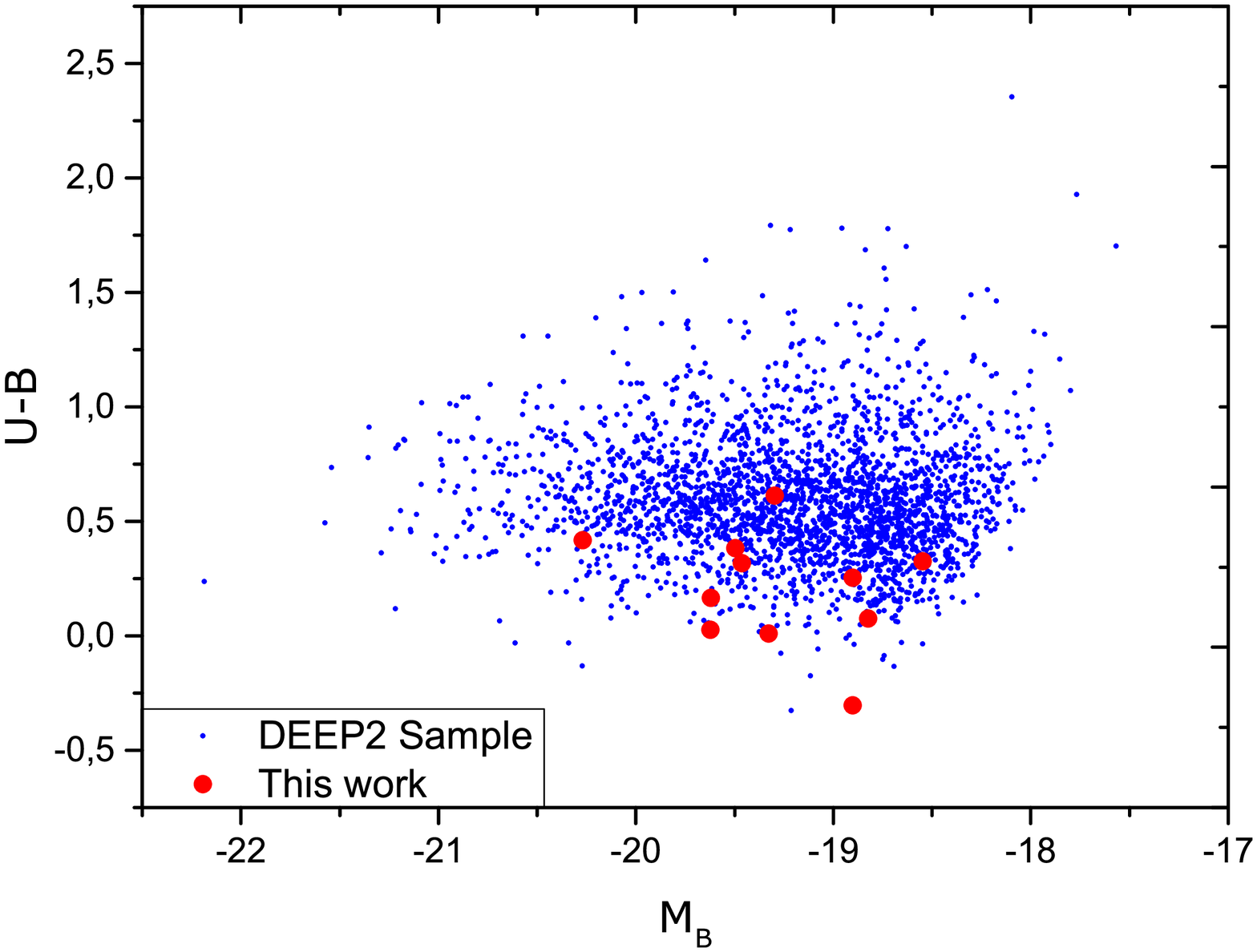}

\includegraphics[scale=0.45]{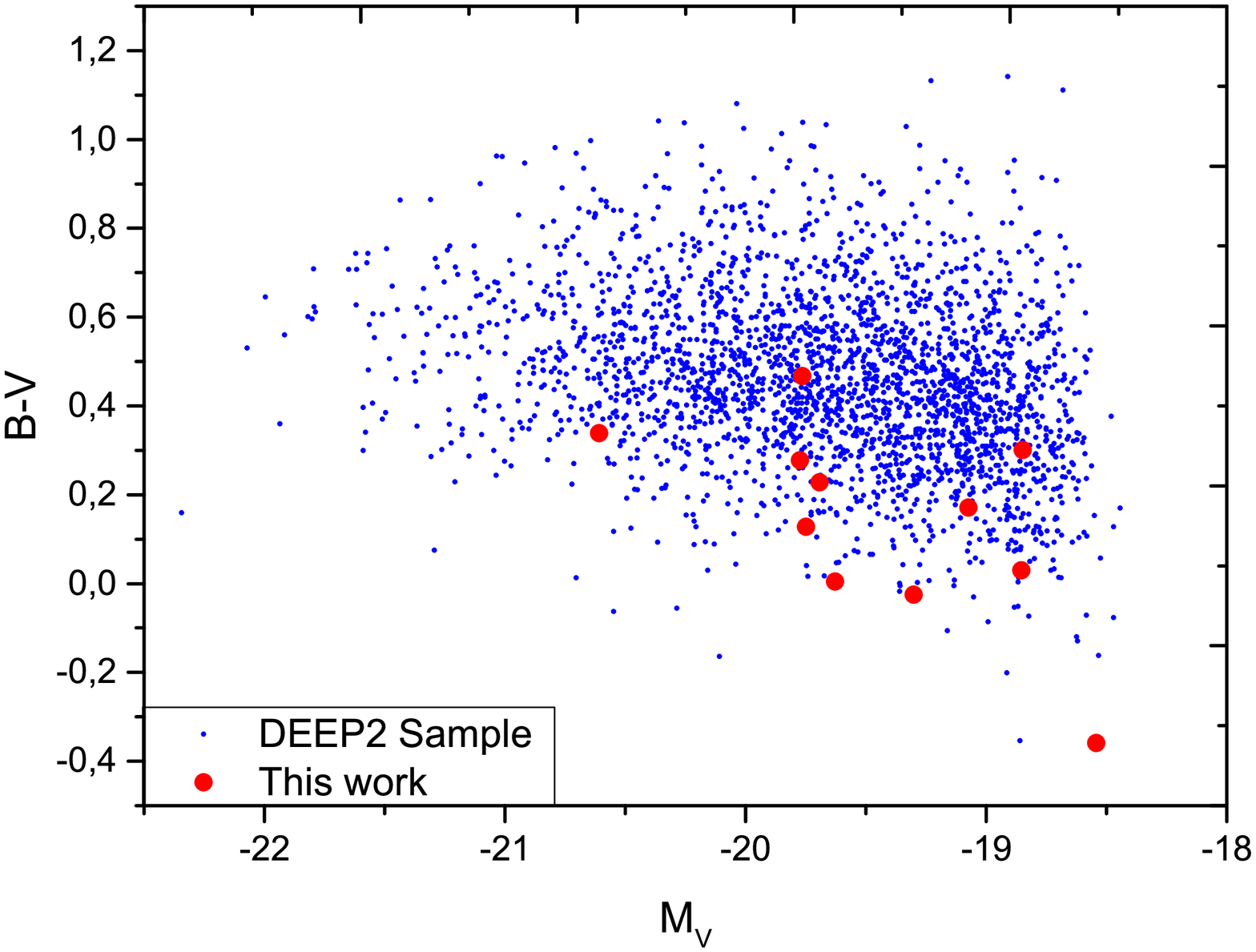}

\caption{Color-Magnitude diagrams for U, B, and V bands. The DEEP2 sample is
composed by 2421 objects in the same interval redshift of this work
and with $EW_{\beta}>10\mathring{A}$. Data obtained from DEEP 2 Automated
Line Fitting.}

\end{center}
\end{figure}

{\large The galaxies of this work were found between the bluer and
more luminous of the sample. These are common properties between star-forming
metal poor galaxies. As we can see in Figure 2, there are not many
objects of this kind for such luminosity and redshift.}{\large \par}

{\large We will analyze the temperatures, electron densities and oxygen
abundances of these galaxies in the following sections.}{\large \par}

\section{{\LARGE Results}}

{\large For short wavelength intervals DEIMOS response is essentially
constant. This allows us to measure some relative line fluxes without
requiring flux calibration of the instrument response. For the analysis
of metallicity we have normalized each oxygen emission line to its
nearest Balmer decrement in order to determine their relative flux
avoiding flux calibrations. }{\large \par}

{\large As H$\text{\textgreek{b}}$ and {[}OIII{]}$\lambda\lambda4959,5007$
are very close in wavelength, the DEIMOS response is almost identical,
so the relative fluxes remain constant with or without calibration.
We can obtain for the same reason the unreddened relative fluxes.
since the galactic extinction law doesn't change too much between
nearby wavelengths. }{\large \par}

{\large However, it is not possible to use this procedure with the
{[}OII{]} $\lambda3727$ doublet since the higher order Balmer decrement
could not be measured accurately. Thus the {[}OII{]} doublet has to
be calibrated and dereddened prior to its measuring as we will see
in the following sections. }{\large \par}

\subsection{{\Large Flux Calibration of {[}OII{]} $\lambda3727$ doublet}}

{\large DEEP2 spectra are not flux calibrated. Hence it is necessary
to perform a relative flux calibration over each spectrum. This kind
of calibration can be made by studying the performance of the CCD
chip to the arrival of photons. We have defined the throughput as
the number of photons that are detected at the CCD versus the number
of photons that leave the reference object. Three methods were used
to measure relative DEIMOS throughput at the DEEP2 survey settings
(Konidaris and Koo. 2003), but the basic algorithm is the same in
all of them. First of all, it was measured a stellar spectrum with
DEIMOS spectrograph, then it has to be determined a ``standard''
spectrum associated with the measured one, that is, what the measured
spectrum should be. Finally the measured spectrum is divided by the
standard one so that the result is called throughput. This can be
approximated by the following fourth order polynomial:}{\large \par}

\begin{equation}
y(\lambda)=-77,9026+0,0395916\lambda-7,49911\times10^{-6}\lambda^{2}+6,2969\times10^{-10}\lambda^{3}-1,97967\times10^{-14}\lambda^{4}
\end{equation}

{\large where $\lambda$ is wavelength in Å and y($\lambda$) is in
units of $throughput/\mathring{A}$. The throughput is applied to
each emission line rate according to the following expression where
N is the number of counts of the fitted emission lines:}{\large \par}

{\large 
\begin{equation}
\frac{F_{1}}{F_{2}}=\frac{N_{1}(\lambda_{1})}{N_{2}(\lambda_{2})}\frac{y(\lambda_{2,obs})}{y(\lambda_{1,obs})}
\end{equation}
}{\large \par}

{\large If two lines are very close in wvalength there is no need
to use this method because the throughput will be roughly the same.
As we will see in the following section, only the {[}OII{]} $\lambda3727$
doublet has been calibrated by this method because there is not any
Balmer line close enough to its wavelength. Konidaris and Koo conclude
that the associated error to this kind of calibration would be at
most 10\%.}{\large \par}

\subsection{{\Large Reddening}}

{\large In the Bohr model of the hydrogen atom there are many different
energy levels between which electrons can transfer if they emit or
absorb the proper amount of energy. Upward moves require absorption
of energy, while downward ones release energy. The Balmer series is
characterized by downward electron transitions from levels above the
second energy level to second level. When many ionized hydrogen atoms
are recombining, as in a planetary nebula where atoms are being ionized
and recombining all the time, the captured electrons cascade down
through the energy levels. emitting photons of the appropriate wavelengths
as they fall. The likelihood of any particular downward jump is dictated
by atomic constants. and thus the ratios of all possible transitions
can be calculated. This leads to the Balmer decrement, the well known
theoretical ratios among the intensities of the Balmer lines.}{\large \par}

{\large Thus, the intensity ratios of Balmer lines in all ionized
regions should be roughly the same. However, this is not what is observed.
Interstellar reddening produced by micron sized dust particles selectively
dims shorter wavelength more than it does longer-wavelength, leading
to Balmer line ratios that differ systematically from the theoretical
predictions. The more dust, the greater the disparity between the
observed and theoretical Balmer decrements. From the size of the discrepancy
between observed and theoretical Balmer decrements, we can infer the
amount of interstellar reddening.}{\large \par}

{\large The amount of extinction is parametrized using the logarithmic
extinction coefficient c(H$\beta$). It was assumed that the major
part of the reddening occurs within the targets objects due to the
position of the observation fields and because the observed wavelength
are fairly red. Consequently the spectra have been dereddened in the
rest frame of the target assuming case B recombination theoretical
line ratios. The logarithmic extinction coefficient c(H$\beta$) has
been derived using standard nebular analysis techniques (Osterbrock.1989)
and assuming the galactic extinction law of Miller \& Mathews (1972)
with $R_{v}$=3.2. The coefficient has been obtained by taking the
difference between the theoretical and observed Balmer decrement available
in the espectra and dividing by the normalized extinction curve so
that $f(H\beta)=1$. We could only use one Balmer emission line, H$\gamma$,
in our sample due to the limited spectral range that the images have
at this redshift: }{\large \par}

{\large 
\begin{equation}
\log\left[\frac{I(\lambda)}{I(H\beta)}\right]=\log\left[\frac{F(\lambda)}{F(H\beta)}\right]+c(H\beta)f(\lambda)
\end{equation}
}{\large \par}

\begin{center}

\begin{table}
\begin{tabular}{cccccccc}
\hline 
\multirow{2}{*}{{\small $ID$}} & \multirow{2}{*}{{\small $z$}} & {\small $f([OII])$} & {\small $f([OIII])$} & {\small $f([OIII])$ } & {\small $f([OIII])$ } & \multirow{2}{*}{{\small $EW(H\lyxmathsym{\textgreek{b}})$}} & \multirow{2}{*}{{\small $c(H\lyxmathsym{\textgreek{b}})$}}\tabularnewline
 &  & $\lyxmathsym{\textgreek{l}}3727$ & {\small $\lyxmathsym{\textgreek{l}}4363$} & {\small $\lyxmathsym{\textgreek{l}}4959$} & {\small $\lyxmathsym{\textgreek{l}}5007$} &  & \tabularnewline
\hline 
\hline 
{\small 13016475} & {\small 0.7468356} & {\small 81.9 \textpm{} 8.2} & {\small 8.4 \textpm{} 0.3} & {\small 209.7 \textpm{} 2.0} & {\small 616.9 \textpm{} 5.8} & {\small 161.5} & {\small 0.48}\tabularnewline
{\small 22032252} & {\small 0.7487180} & {\small 134.6 \textpm{} 13.7} & {\small 10.3 \textpm{} 1.0} & {\small 159.5 \textpm{} 2.4} & {\small 452.2 \textpm{} 6.7} & {\small 78.0} & {\small 0.45}\tabularnewline
{\small 31019555} & {\small 0.7552315} & {\small 70.9 \textpm{} 7.2} & {\small 9.5 \textpm{} 0.5} & {\small 176.7 \textpm{} 2.3} & {\small 502.2 \textpm{} 5.4} & {\small 164.9} & {\small 0.08}\tabularnewline
{\small 12012181} & {\small 0.7716637} & {\small 182.3 \textpm{} 18.5} & {\small 14.4 \textpm{} 1.1} & {\small 208.8 \textpm{} 3.0} & {\small 630.7 \textpm{} 8.0} & {\small 41.7} & {\small 1.23}\tabularnewline
{\small 14018918} & {\small 0.7709119} & {\small 187.5 \textpm{} 18.8} & {\small 6.2 \textpm{} 0.6} & {\small 190.0 \textpm{} 1.5} & {\small 578.8 \textpm{} 5.7} & {\small 124.1} & {\small 0.78}\tabularnewline
{\small 41059446} & {\small 0.7743866} & {\small 157.3 \textpm{} 15.9} & {\small 7.7 \textpm{} 1.0} & {\small 167.0 \textpm{} 2.0} & {\small 411.3 \textpm{} 5.7} & {\small 34.8} & {\small 0.00}\tabularnewline
{\small 41006773} & {\small 0.7838417} & {\small 171.9 \textpm{} 17.7} & {\small 12.6 \textpm{} 1.4} & {\small 168.6 \textpm{} 3.5} & {\small 515.9 \textpm{} 10.2} & {\small 36.0} & {\small 0.98}\tabularnewline
{\small 22021909} & {\small 0.7979995} & {\small 133.8 \textpm{} 13.5} & {\small 13.3 \textpm{} 0.7} & {\small 197.6 \textpm{} 3.2} & {\small 598.6 \textpm{} 7.1} & {\small 25.2} & {\small 0.24}\tabularnewline
{\small 22020856} & {\small 0.7944890} & {\small 200.6 \textpm{} 20.2} & {\small 8.0 \textpm{} 1.0} & {\small 160.1 \textpm{} 2.1} & {\small 450.6 \textpm{} 5.3} & {\small 67.7} & {\small 0.73}\tabularnewline
{\small 22020749} & {\small 0.7967923} & {\small 217.3 \textpm{} 22} & {\small 9.7 \textpm{} 1.2} & {\small 125.3 \textpm{} 2.6} & {\small 403.8 \textpm{} 6.6} & {\small 102.3} & {\small 0.49}\tabularnewline
{\small 31046514} & {\small 0.7885643} & {\small 188.9 \textpm{} 19.2} & {\small 6.0 \textpm{} 0.7} & {\small 169.5 \textpm{} 3.0} & {\small 480.2 \textpm{} 7.3} & {\small 47.0} & {\small 0.85}\tabularnewline
\hline 
\end{tabular}

\caption{Oxygen emission line fuxes in counts normalized to $f(H\text{\textgreek{b}})=100$.
Rest frame equivalent width of$(H\lyxmathsym{\textgreek{b}})$ and
logarithmic extinction coefficient computed from the $[OII]$ $\lambda3727$
doublet}
\end{table}
\end{center}

\subsection{{\Large Temperatures and Densities}}

{\large To compute chemical abundances in ionized gas nebulae, it
is required to know the electron temperature. If there were not temperature
gradients we could think in an isothermal distribution of the temperature.
However the temperature gradients throughout the gas region force
us to use the appropiate line temperature for the calculation of each
ion. These temperatures can be deduced from the corresponding emission
line ratio, but usually not all the lines are accessible in the spectra
or have large errors. In these cases, some assumptions are usually
adopted concerning the temperaure structure through the nebula. In
this work we assume a two phase model with a low ionization zone which
depens emission of the {[}OII{]} $\lambda3727$ doublet, and a high
ionization zone in which the {[}OIII{]} lines are formed.}{\large \par}

{\large Now we can determine the values of electron density and temperature
from the flux ratios of the observed oxygen emission lines. These
ratios are summarized in Table 3.}{\large \par}

\begin{center}

\begin{table}
\begin{centering}
\begin{tabular}{cl}
\hline 
\multicolumn{2}{c}{Ratios}\tabularnewline
\hline 
$t_{e}[OIII]$ & $R_{O3}=[I(4959)+I(5007)]/I(4363)$\tabularnewline
$t_{e}[OII]$ & $R_{O2}=I(3727)/[I(7319)+I(7330)]$\tabularnewline
$n_{e}[OII]$ & $R_{n_{e}[OII]}=I(3726)/I(3729)$\tabularnewline
\hline 
\end{tabular}
\par\end{centering}

\caption{Emission-line ratios used to derive electron densities and temperatures}

\end{table}

\end{center}

{\large As we said before the ratio which involved the 3727 doublet
have the greatest associated error since it had to be calibrated and
reddening corrected. We have derived the physical conditions of the
ionized gas using the expressions given by Hägele et al. 2008 for
the oxygen emission lines of HII galaxies. These formulas are approximate
expressions for the statistical equilibrium model in a five levels
atom. We present below the adequate fitting functions they used from
the TEMDEN task of IRAF, which is based on the program FIVEL (De Robertis,
Dufour \& Hunt 1987; Shaw \& Dufour 1995):}{\large \par}

{\large 
\begin{equation}
t_{e}([OIII])=0.8254-0.0002415R_{O3}+\frac{47.77}{R_{O3}}
\end{equation}
}{\large \par}

{\large 
\begin{equation}
t_{e}([OII])=0.23+0.0017R_{O2}+\frac{38.3}{R_{O2}}+f_{1}(n_{e})
\end{equation}
}{\large \par}

{\large Since we do not have all the oxygen lines in the allowed spectral
range, e.g. $[OII]\lambda\lambda7319,7330$, it was necessary to calculate
$t_{e}[OII]$ from $t_{e}[OIII]$ using Stasinska models as presented
in Pagel et al. (1992): }{\large \par}

{\large 
\begin{equation}
t_{e}([OII])\text{\textsuperscript{-}\textonesuperior}=0.5(t_{e}[OIII]\text{\textsuperscript{-}\textonesuperior}+0.8)
\end{equation}
}{\large \par}

{\large This expression is the result of an experimental fitting over
the theoretical models proposed by Stasinka in 1990. We are treating
with star forming galaxies in which we expect to find high temperature
ionized regions. Typically the {[}OIII{]} electron temperature is
expected to be greater than the {[}OII{]} temperature since the first
comes from the hotter part of the star-forming region while the second
come from the colder regions. Errors in the measurements of $t_{e}[OII]$
are normally higher than the $t_{e}[OIII]$ ones due to the electronic
density dependence of the {[}OII{]} lines. However, the $t_{e}[OII]$
erros in our sample will be lower considering that we are using the
relation between both electron temperatures through the Stasinska
model so $t_{e}[OII]$ depends exclusively of the $t_{e}[OIII]$ value.
This assumption can or can not be a good approximation depending of
the density nebular conditions. Since the assumed state of these regions
agrees with the case B recombination phase with temperature around
$10\text{\textsuperscript{4}}K$ and density around $100cm\text{\textsuperscript{-}\textthreesuperior}$,
we can neglect the density effects over $t_{e}[OII]$ and rely on
the Stasinska model. }{\large \par}

{\large The derivation of $R_{O3}$ ratio is more complex than in
other cases because it involves lines that are widely separate through
the spectrum, and therefore they have different response function
from DEIMOS spectrograph. As mentioned in section 2, the relative
flux of each oxygen line has been measured relative to its nearest
Balmer decrement, so that for $[OIII]\lambda4363$ we measure respect
to $H\gamma$ while for the relative flux measurements of {[}OIII{]}$\lambda\lambda4959,5007$
lines we used $H\beta$. To obtain the correct $R_{O3}$ratio we assume
the theoretical value between the used Balmer lines obtaining $H\gamma/H\beta=0.471$
(assuming case B recombination). This way we obtain an expression
for the $R_{O3}$ ratio minimizing the associated errors of the DEIMOS
flux calibration $H\beta$:}{\large \par}

{\large 
\[
R_{O3}=\frac{\frac{f(4959)+f(5007)}{f(H\beta)}}{\frac{f(4363)}{f(H\gamma)}}\frac{I(H\gamma)}{I(H\beta)}
\]
}{\large \par}

{\large This method allow us to compute the electronic temperature
with deviations not bigger than 5\% in the most of cases. Table 4
shows the results for the electron density and temperature. The objects
with $t_{e}[OII]\sim t_{e}[OIII]$ are two exceptions since their
ionization structure differs from the expected for this type of objects.
The difference between the values of $t_{e}([OIII])$ and $t_{e}([OII])$
is slight enough to be overlapped by the predicted errors. The density
of the sample remains in its expected value around $100cm\text{\textsuperscript{-}\textthreesuperior}$,
far away from the critical oxygen density values, which confirms our
assumptions about Stasinska model for this type of objects.}{\large \par}

\begin{table}
\begin{center}

\begin{tabular}{ccccc}
\hline 
{\small $ID$} & $R_{n_{e}[OII]}$ & {\small $n_{e}([OII])$} & {\small $t_{e}([OIII])$} & {\small $t_{e}([OII])$}\tabularnewline
\hline 
\hline 
{\small 13016475} & 0.849 & {\small 202$\pm$52} & {\small 1.285 \textpm{} 0.020} & {\small 1.267 \textpm{} 0.010}\tabularnewline
{\small 22032252} & 0.711 & $\leq$50 & {\small 1.612 \textpm{} 0.084} & {\small 1.408 \textpm{} 0.032}\tabularnewline
{\small 31019555} & 0.863 & {\small 225$\pm$50} & {\small 1.475 \textpm{} 0.040} & {\small 1.353 \textpm{} 0.017}\tabularnewline
{\small 12012181} & 0.734 & {\small 70} & {\small 1.632 \textpm{} 0.066} & {\small 1.416 \textpm{} 0.025}\tabularnewline
{\small 14018918} & 0.835 & {\small 183$\pm$78} & {\small 1.179 \textpm{} 0.042} & {\small 1.213 \textpm{} 0.022}\tabularnewline
{\small 41059446} & 0.768 & {\small 109$\pm$53} & {\small 1.444 \textpm{} 0.086} & {\small 1.340 \textpm{} 0.037}\tabularnewline
{\small 41006773} & 0.794 & {\small 142$\pm$75} & {\small 1.693 \textpm{} 0.104} & {\small 1.438 \textpm{} 0.038}\tabularnewline
{\small 22021909} & 0.687 & $\leq$50 & {\small 1.609 \textpm{} 0.050} & {\small 1.407 \textpm{} 0.019}\tabularnewline
{\small 22020856} & 0.793 & {\small 138$\pm$72} & {\small 1.435 \textpm{} 0.078} & {\small 1.336 \textpm{} 0.034}\tabularnewline
{\small 22020749} & 0.743 & {\small 81$\pm$53} & {\small 1.691 \textpm{} 0.116} & {\small 1.438 \textpm{} 0.042}\tabularnewline
{\small 31046514} & 0.740 & {\small 77$\pm$42} & {\small 1.239 \textpm{} 0.054} & {\small 1.245 \textpm{} 0.027}\tabularnewline
\hline 
\end{tabular}

\caption{Oxygen temperatures and densities for the galaxy sample. The temperature
is given in units of $10\text{\textsuperscript{4}}K$ while the density
is in $cm\text{\textsuperscript{-}\textthreesuperior}$.}

\end{center}
\end{table}

\subsection{{\Large Chemical Abundances}}

{\large Abundances refer to the relative proportions of atomic species
in the gas phase nebula and these are usually expressed relative to
Hydrogen. }{\large \par}

{\large The stimation of the metallicity of the emitting nebula is
one of the main goals of any nebular analysis. The metallicity of
the ionized gas in a galaxy allows us to know the updated chemical
composition of the most recent generation of the stars born in the
galaxy studied (C. Hoyos 2006). Oxygen is particularly important considering
that ionic species of O found in nebulae have strong optical and near
ultraviolet emission lines, and it acts as a coolant in most nebulae.}{\large \par}

{\large In order to determine the oxygen ionic abundances it was necessary
the measurement of two emission line ratios, $\mbox{[OII]}\lambda\lambda3726+3729/H\beta$
and {[}OIII{]} $\lambda\lambda4959+5007/H\beta$. We used again the
fitted expressions given by Hägele et al. 2008 for the oxygen emission
lines in HII galaxies in order to compute the ionic abundances. In
this case they fitted the IONIC task of IRAF which is based on the
program FIVEL (De Robertis, Dufour \& Hunt 1987; Shaw \& Dufour 1995): }{\large \par}

\begin{equation}
12+log(O^{+}/H^{+})=log\frac{I(3726+3729)}{I(H\beta)}+5.992+\frac{1.583}{t_{e}[OII]}-0,681log(t_{e}[OII])+log(1+2.3n_{e})
\end{equation}

\begin{equation}
12+log(O^{++}/H^{+})=log\frac{I(4959+5007)}{I(H\beta)}+6.144+\frac{1.251}{t_{e}[OIII]}-0.550log(t_{e}[OIII])
\end{equation}

{\large The main problem with the abundance results resides in the
method used to get the {[}OII{]} temperature. As we stated before
our spectral range does not allow us to observe all the necessary
{[}OII{]} lines to compute $t_{e}[OII].$ That is the reason why we
used the Stasinska method, with the assumption of an ionization structure
composed of only two phases ({[}OIII{]} and {[}OII{]}) for these objects.}{\large \par}

{\large The metallicity of the galaxies is one of the most important
properties to measure if we want to understand their evolutionary
behaviour. THe total oxygen abundance is a common way to compute the
metallicity of metal-poor galaxies, but prior to measure the total
oxygen abundance we have to measure the abundances of all its ions.
Since the most abundant ions of oxygen in this regions are $O^{+}$and
$O^{2+}$, we can determine the oxygen total abundance as:}{\large \par}

{\large 
\begin{equation}
\frac{O}{H}=\left(\frac{O^{+}}{H^{+}}\right)+\left(\frac{O^{2+}}{H^{+}}\right)
\end{equation}
}{\large \par}

{\large In Tabla \ref{Tabla 5} we provide the resultant oxygen ionic
and total abundances and the ratio between oxygen ionization states
($\log(O\text{\textsuperscript{+}\textsuperscript{+}\ensuremath{/O{{}^+}})}$
in the nebula. The errors are smaller than 0.1 dex for the oxygen
ionic and total abundances and between $0.1$ and $0.15$ dex for
the $[OIII]/[OII]$ ratio. The results obtained shows low oxygen abundances
with metallicities fron 1/3 to 1/10 of the solar value. These results
together with the color-magnitude diagrams allow us to confirm that
our galaxy sample is indeeed formed by metal-poor galaxies.}{\large \par}

\begin{table}
\begin{center}

\begin{tabular}{ccccc}
\hline 
{\small $ID$} & $12+log(O\text{\textsuperscript{+}}/H\text{\textsuperscript{+}})$ & $12+log(O\text{\textsuperscript{+}\textsuperscript{+}}/H\text{\textsuperscript{+}})$ & $12+log(O/H)$ & $log(O{{}^+}{{}^+}/O{{}^+})$\tabularnewline
\hline 
\hline 
{\small 13016475} & 7.08 \textpm{} 0.06 & 7.97 \textpm{} 0.02 & 8.03 \textpm{} 0.02 & 0.89 \textpm{} 0.08\tabularnewline
{\small 22032252} & 7.14 \textpm{} 0.08 & 7.6 \textpm{} 0.06 & 7.73 \textpm{} 0.06 & 0.46 \textpm{} 0.13\tabularnewline
{\small 31019555} & 6.92 \textpm{} 0.06 & 7.74 \textpm{} 0.03 & 7.80 \textpm{} 0.04 & 0.82 \textpm{} 0.09\tabularnewline
{\small 12012181} & 7.27 \textpm{} 0.07 & 7.71 \textpm{} 0.04 & 7.84 \textpm{} 0.05 & 0.44 \textpm{} 0.11\tabularnewline
{\small 14018918} & 7.51 \textpm{} 0.07 & 8.04 \textpm{} 0.05 & 8.15 \textpm{} 0.05 & 0.53 \textpm{} 0.12\tabularnewline
{\small 41059446} & 7.28 \textpm{} 0.08 & 7.74 \textpm{} 0.07 & 7.87 \textpm{} 0.07 & 0.46 \textpm{} 0.15\tabularnewline
{\small 41006773} & 7.22 \textpm{} 0.08 & 7.58 \textpm{} 0.06 & 7.74 \textpm{} 0.07 & 0.36 \textpm{} 0.14\tabularnewline
{\small 22021909} & 7.14 \textpm{} 0.06 & 7.7 \textpm{} 0.03 & 7.81 \textpm{} 0.04 & 0.56 \textpm{} 0.10\tabularnewline
{\small 22020856} & 7.39 \textpm{} 0.08 & 7.73 \textpm{} 0.06 & 7.89 \textpm{} 0.07 & 0.34 \textpm{} 0.14\tabularnewline
{\small 22020749} & 7.32 \textpm{} 0.08 & 7.45 \textpm{} 0.07 & 7.69 \textpm{} 0.08 & 0.13 \textpm{} 0.16\tabularnewline
{\small 31046514} & 7.48 \textpm{} 0.08 & 7.93 \textpm{} 0.06 & 8.06 \textpm{} 0.06 & 0.45 \textpm{} 0.14\tabularnewline
\hline 
\end{tabular}

\caption{Oxygen ionic and total abundances of the selected galaxy sample in
section 2. We enclose as well the ionic abundance ratio which denotes
the prevalence of the {[}OIII{]} ionic state in the gas region.}
\label{Tabla 5}

\end{center}
\end{table}

\section{{\LARGE Discussion}}

\begin{figure}[H]
\begin{centering}
\includegraphics[scale=0.5]{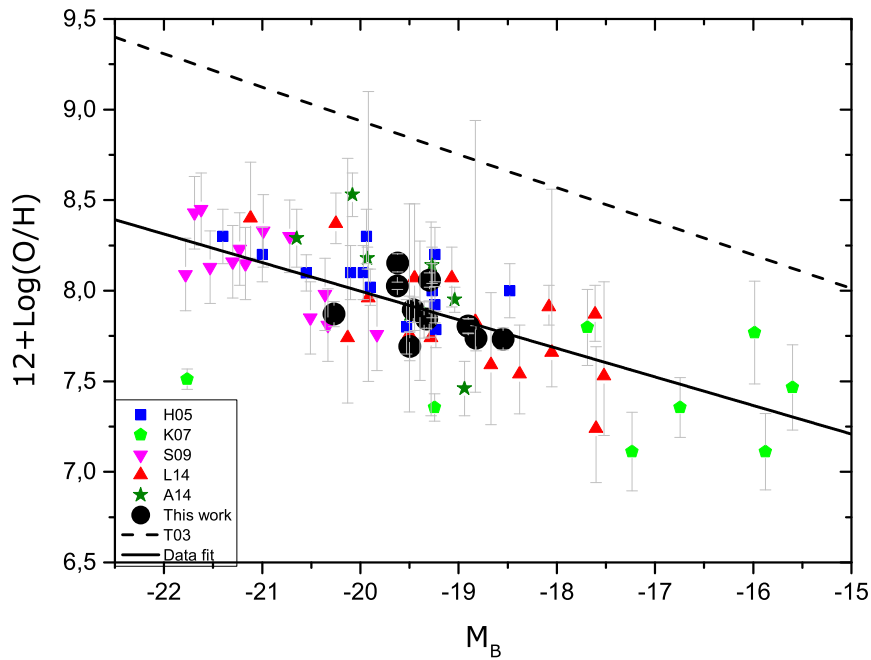}
\par\end{centering}

\caption{The Luminosity-Metallicity Diagram for intermediate star-forming galaxies
for objects of this study and comparable samples from the literature.
H05 includes 15 luminous star-forming galaxies from DEEP2. K07 includes
8 EELGs 6 of them are extremely low metallicity galaxies. S09 includes
13 metal-poor galaxies. A14 includes 7 EELGs. L14 includes 17 metal-poor
objects. This work presents 11 new metal-poor galaxies. The dashed
line on the upper zone of the diagrama represents the Tremonti luminosity-metallicity
relation for local SDSS galaxies (Tremonti et al. 2004, T04 in the
diagram) while the solid line represents the best fit to all the objects
collected. }

\label{met_lum}
\end{figure}

{\large The main result of this work can be seen in Figure \ref{met_lum}
where we compare different sets of metal-poor galaxies in a metallicity-luminosity
diagram. This figure shows the total oxygen abundance for the 11 galaxies
in our sample which is between 1/3 and 1/10 of the solar value of
$12+\log[O/H]_{\odot}$=8.69 (Allende Prieto et al. 2001). We compare
our sample with other similar ones: Hoyos et al. (2005, hereafter
H05), Kakuzu et al. (2007, hereafter K07), Salzer et al. (2009, hereafter
S09), Amorín et al. (2014, hereafter A14) and Ly et al. (2014, here
after L14) which, as far as I know, represent the almost total comparable
information in the literature. It can be seen that our data are in
good agreement with these previous works. }{\large \par}

{\large The sample from A14 is composed by extreme emission line galaxies
(EELGs) selected from the 20k zCOSMOS Bright Survey by their unusually
large {[}OIII{]} $\lambda5007$ equivalente widths. They are seven
purely star-forming galaxies with the {[}OIII{]} $\lambda4363$ line
in their spectrawith redshifts from $z=0.43$ to $0.63$.}{\large \par}

{\large The sample from K07 was collected from spectroscopic observations
of 161 Ultra strong emission line galaxies (USELs) using the DEIMOS
spectrograph on the Keck II telescope. The galaxies are spread in
a wide range of redshifts ($0.38<z<0.83$) but is clearly biased towards
low metallicities (in some cases with values below that of I Zw 18)
and the major part of the sample has low luminosity values. Within
this sample are found some objects with quoted electron temperatures
higher than 20000K and up to $30000K$ which can not be explained
as having normal hiot stars as the only source of ionization. No diagnostic
diagrams are provided in order to decide if these objects show any
contribution by an AGN or shock heating. If this contribution were
important, the nature of the galaxy emission lines in K07 sample would
be contaminated and different to the others presented in Figure \ref{met_lum}.
As a conservative approach, these data should be excluded of the luminosity-metallicity
relation. We suggest to reobserve these objects in future works.}{\large \par}

{\large The sample from S09 presents some of the most luminous objects
of this type ($-22<M_{B}<-20$) at intermediate redshifts ($0.35<z<0.41$)
and cover a wide range of the oxygen metallicty of the entire sample.
The star-forming galaxies of the KISS sample are derived from a wide-field
Schmidt survey that selects emission line objects via the presence
of $H\alpha$ emission in their objective-prism spectra (Salzer et
al. 2000). The selection process includes a filter that restricted
the wavelength coverage of the slitless spectra to $6400-7200$$\AA$.
As we move towards higher redshift the bandpass allows to detect objects
with other strong emission lines such as {[}OIII{]}$\lambda5007$
due to the shift of its wavelength. Some of the objects selected by
this criteria are presented in the left area of Figure \ref{met_lum}.}{\large \par}

{\large The sample from L14 encompasses a wide luminosity range (from
$M_{B}=-21,1$ to $-17.5)$ with abundances going from extreme metal
poor galaxies ($12+\log(O/H)\lesssim7,65$) to galaxies with almost
solar abundances. In addition the redshift dispersion of the sample
is similar to the one shown in K07$(0.38<z<0.83)$ what provides a
good chance to observe star-forming galaxy properties at different
ages. The data has been gathered using optical spectroscopy with DEIMOS
and MMT's hectospec spectrographs.}{\large \par}

{\large The sample from H05 and this work have many features in common
since both of them have been taken from the DEEP2 redshift survey
with the DEIMOS spectrograph. Both samples cover the central region
of the diagram and have smaller dispersion and smaller errors than
the others. The accuracy of this work has been improved respect to
H05 results thanks to the measurement of the {[}OII{]}$\lambda\lambda3726,3729$
doublet, which allows to compute directly the nebular electron density
and the O\textsuperscript{+}/H\textsuperscript{+} ionic abundance
with error bars below 0.1 dex as we can see in Figure \ref{met_lum}.
The redshift range of the H05 sample goes from $z=0.51$ to $0.85$
while in this work the redshift is more restricted ($0.744<z<0.835)$
due to the {[}OII{]} line detection requirements. }{\large \par}

{\large It can be seen from the figure that both relations show an
offset between them. The fit for the Tremonti L-Z relation, for SDSS
galaxies with a redshift distribution around $z\sim0.08,$ and the
one obtained in this work for the various collected samples of metal-poor
galaxies at intermediate redshift $(0.4<z<0.8),$are respectively
given below:}{\large \par}

\begin{equation}
12+\log(O/H)=-0.185(\pm0.001)M_{B}+5.238(\pm0.018)
\end{equation}

\begin{equation}
12+\log(O/H)=-0.158(\pm0.020)M_{B}+4.842(\pm0.381)
\end{equation}

{\large Our results seem to follow a linear relation between logarithmic
oxygen abundance and absolute magnitude with a slope similar to that
of T04 but with an offset of about 1.0 dex in abundance, wich can
be due to the different nature of the objects involved, selection
effects regarding both luminosity and cehmical abundances, or a genuine
evolutionary effect. Further work is needed to obtained a Mass-Metallicity
relation which can be more free of selection effects. }{\large \par}

\begin{figure}
\begin{centering}
\includegraphics[scale=0.45]{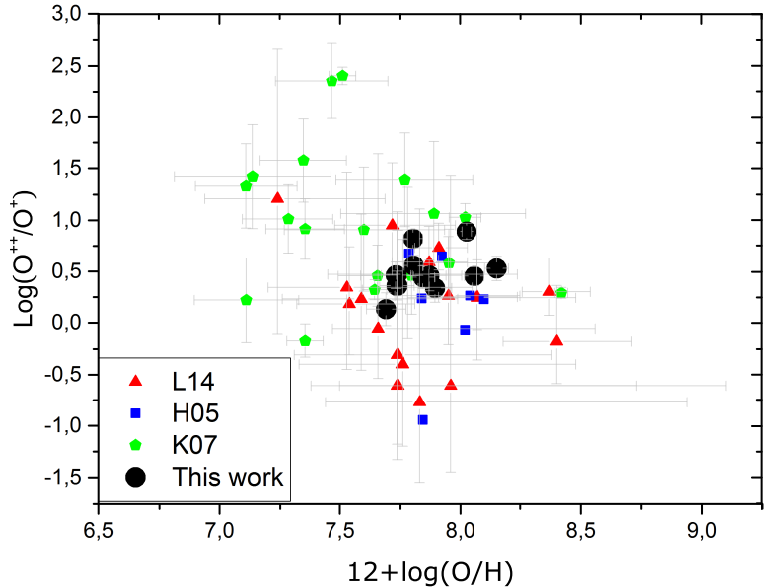}
\par\end{centering}

\caption{Oxygen ionic fraction as a function of metallicity for the three galaxy
samples for wich this information can be derived, together with the
objects of this work. A general trend of decreasing ionization degree
with increasing metallicity is present along with a strong degeneracy
effect in the metallicity with large scatter. }

\label{o+++}
\end{figure}

{\large A hint on the similar or different nature of the objects in
our various considered samples can be obtained by looking at the oxygen
ionization fractions which reflect the ionization structure of the
nebulae. Figure \ref{o+++} shows the logarithmic ratio between the
O\texttwosuperior{}\textsuperscript{+} and O\textsuperscript{+} ionic
abundances as a function of the total oxygen abundance. A general
trend of decreasing ionization degree (lower O\texttwosuperior{}\textsuperscript{+}/O\textsuperscript{+}
fraction) with increasing metallicity is expected from the cooling
of the gas. This trend is present in the figure, although with a scatter
which is larger tahn observational errors. Some degeneracy is found
with objects of similar metallicity and widely different ionization
conditions. Again further work and more data are needed in order to
explore these effects. }{\large \par}

{\large The logarithmic ratio of the {[}OIII{]}/{[}OII{]} lines is
often used as a functional parameter to represent the degree of nebular
excitation and can also be used to characterised the different samples
involved. Figure 5 shows this ratio ($O_{32}=([OIII]\lambda4959+\lambda5007)/[OII]\lambda3727$)
as a function of logarithmic oxygen abundance. Taking figures 4 and
5 together, we can see that a relation between the $O_{32}$ parameter
an the oxygen ionic fraction is questionable. In particular, the fact
that all the the K07 objects show a very similar value of $O_{32}$
(between 0 and 0.4) spanning, in contrast, a wide range of metallicities.
This results together with the mentioned too high temperatures of
many objects of the K07 sample reinforce our assumption of an alternative
heating process for this objects, since a contribution by shocks will
enhance the {[}OII{]} emission thus providing an artificially low
values of $O_{32}$.}{\large \par}

\begin{figure}
\begin{centering}
\includegraphics[scale=0.5]{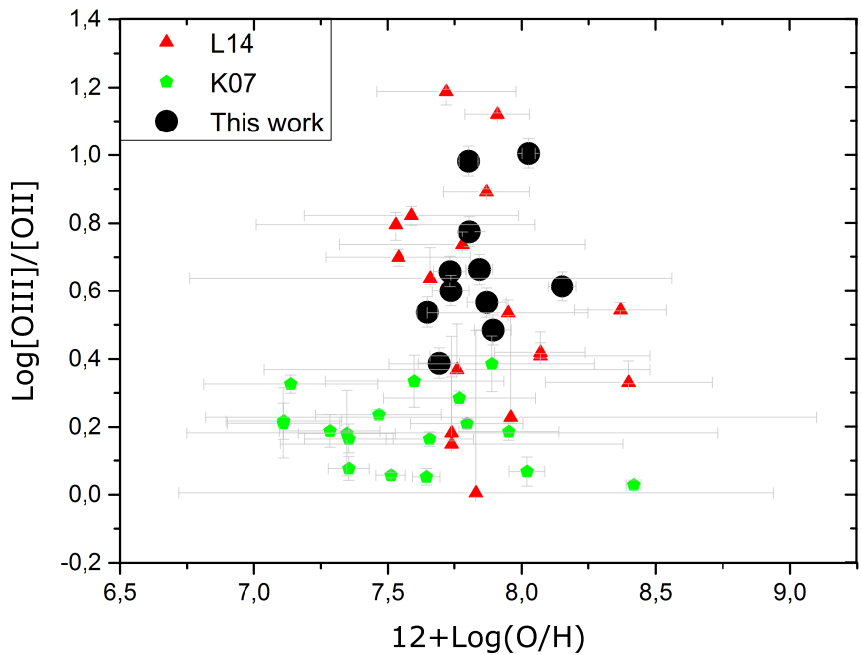}
\par\end{centering}

\caption{Logarithmic $O_{32}$ emission line ratio as a function of metallicity
for the three galaxy samples for which this ratio can be computed. }
\end{figure}

\section{{\LARGE Conclusions}}

{\large We have studied 11 luminous, blue, and metal poor (1/3 to
1/10 of the solar metallicity) galaxies from DEEP2 survey at redshift
$0.744<z<0.835$. The sample was selected by the presence of the {[}OIII{]}
$\lambda4363$ auroral line and the {[}OII{]} $\lambda\lambda3726,3729$
doublet together with the strong {[}OIII{]} emission lines. These
objects represent the $0.002\%$ of the total galaxies within the
same redshift interval in the DEEP2 survey what means that these objects
are rare, although they are of special relevance to understand the
processes of galaxy evolution. }{\large \par}

{\large The computed total oxygen abundances $($between $7.69$ and
$8.15)$ do not differ so much from those published by Hoyos et al.
(2005), but these works together with other such as Ly et al. 2014
constitute the only available and comparable data for objects of this
type. In the Luminosity-Metallicty diagram, galaxies of intermediate
redshift define a linear relation between logarithmic oxygen abundance
and absolute magnitude, arise a linear relation which is almost pararell
to the one proposed by Tremonti et al. (2004) for SDSS local objects,
but is offset by 1 dex to lower metallicity values. }{\large \par}

{\large The ionization structure of the analyzed samples show a general
trend of decreasing ionization degree (lower O\texttwosuperior{}\textsuperscript{+}/O\textsuperscript{+}
fraction) with increasing metallicity which is expected from the cooling
of the gas. This trend can be seen in Figure \ref{o+++}, although
with a larger scatter than the observational errors. Taking the results
of this work together with those presented in L14 and K07, we conclude
that the relation between the }$O_{32}$ {\large parameter and the
oxygen ionic fraction is questionable for metal-poor galaxies since
we observed degeneracy effects between objects of similar metallicity
and widely different ionization conditions.}{\large \par}

{\large Regardless of the cause, metal-poor star-forming galaxies
are highly interesting objects whose study is indispensable to understand
tke processes of galaxy evolution in time.}{\large \par}

\end{document}